\documentclass[aps,twocolumn,prl,superscriptaddress,amsmath,amssymb]{revtex4-1}

\usepackage{times}
\usepackage{amsmath}
\usepackage{amsfonts}
\usepackage{amssymb}
\usepackage{dcolumn}
\usepackage{bm}
\usepackage{txfonts}
\usepackage{ulem}
\usepackage{braket}
\usepackage[dvipdfmx]{graphicx,hyperref}
\usepackage[usenames]{xcolor}

\def\journal #1#2#3#4{{#1} {\bf #2}, #3 (#4).}

\begin{document}
\title{Band Crossings in Honeycomb-layered Transition Metal Compounds}

\author{Yusuke Sugita}
\affiliation{Department of Applied Physics, The University of Tokyo, Bunkyo, Tokyo 113-8656, Japan}
\author{Yukitoshi Motome}
\affiliation{Department of Applied Physics, The University of Tokyo, Bunkyo, Tokyo 113-8656, Japan}

\date{\today}
\begin{abstract}
Two-dimensional electron dispersions with peculiar band crossings provide a platform for realizing topological phases of matter. Here we theoretically show that the $e_g$-orbital manifold of honeycomb-layered transition metal compounds accommodates a plethora of peculiar band crossings, such as multiple Dirac point nodes, quadratic band crossings, and line nodes. 
From the tight-binding analysis, we find that the band topology is systematically changed by the orbital dependent transfer integrals on the honeycomb network of edge-sharing octahedra, which can be modulated by distortions of the octahedra as well as chemical substitutions.
The band crossings are gapped out by the spin-orbit coupling, which brings about a variety of topological phases distinguished by the spin Chern numbers.
The results provide a comprehensive understanding of the previous studies on various honeycomb compounds.
We also propose another candidate materials by {\it ab initio} calculations.
\end{abstract}
\maketitle

%%%%% introduction %%%%%
Two-dimensional materials with layered structure have attracted considerable attention as a good playground for topological states of matter. 
The representative example is monolayer graphene composed of a purely two-dimensional honeycomb network of carbon atoms~\cite{Novoselov666}.
The low-energy excitation in graphene is governed by $\pi$ electrons in $2p$ orbitals, whose energy dispersion has linear band crossings with the Dirac point nodes (DPNs) at the Fermi level, called the Dirac cones.  
Stimulated by the theoretical proposal that the Dirac electron system is potentially changed into a $\mathbb{Z}_2$ topological insulator by the relativistic spin-orbit coupling (SOC)~\cite{PhysRevLett.95.146802,PhysRevLett.95.226801}, graphene and similar honeycomb-monolayer forms of Si and Ge have been studied~\cite{PhysRevB.75.041401,PhysRevLett.107.076802}.  
In addition, few-layer graphene has also received attention as the low-energy spectrum takes a peculiar form depending on the stacking manner.  
For instance, in a bilayer system with the so-called AB stacking, the DPNs turn into quadratic band crossings (QBCs).
As the QBCs possess an instability toward a quantum anomalous (spin) Hall state~\cite{PhysRevLett.103.046811,PhysRevB.89.201110}, the effect of electron correlations has been intensively studied in bilayer graphene~\cite{PhysRevB.81.041401,PhysRevB.81.041402,PhysRevLett.104.156803,PhysRevLett.117.086404}.

Recently, transition metal (TM) compounds with similar honeycomb-layered structure have gained great interest from the peculiar band topology in their $d$-orbital manifolds.
For instance, the DPNs, QBCs, and topological phases were found in the systems with corner-sharing network of the octahedral ligands, e.g., [111] layers of the perovskite structure~\cite{xiao2011interface,PhysRevB.84.201103,PhysRevB.84.201104,PhysRevB.85.245131,PhysRevB.93.165145}, and with edge-sharing octahedra, e.g., trichalcogenides~\cite{PhysRevB.97.035125}, trihalides~\cite{he2016unusual,C6NR08522A,PhysRevB.95.045113,PhysRevB.95.201402}, corundum~\cite{PhysRevB.92.235102,PhysRevB.97.035126}, and rhombohedral materials~\cite{PhysRevB.92.125109,PhysRevB.94.125134,kim2017realizing}.
Interestingly, the number and position of the DPNs as well as the shape of the Dirac dispersions strongly depend on the materials.
This suggests the controllability of the band crossings and topological phases by using the $d$-orbital degrees of freedom, but such an interesting possibility has not been investigated systematically.

In this Rapid Communication, 
we theoretically show that the $e_g$-orbital systems with edge-sharing octahedra can host a plethora of peculiar band crossings and associated topological phases of matter. 
Analyzing a tight-binding model for the $e_g$ manifold, we find that a variety of band crossings appears at half filling of the $e_g$ electrons, such as multiple DPNs, QBCs, and line nodes.
We find that the band topology changes systematically for the orbital dependent transfer integrals, which can be controlled by distortions of the ligand octahedra as well as chemical substitutions. 
We also show that the SOC turns the electronic states with the different band crossings into different types of topological phases characterized by the spin Chern numbers, some of which are unusually high or $\mathbb{Z}_2$ nontrivial.
Our results provide a systematic understanding of the existing {\it ab initio} studies for the honeycomb-layered TM compounds.
Furthermore, we propose candidate materials by using {\it ab initio} calculations, which potentially realize the wide variety of the peculiar band topology.

%%%%% tight-binding model %%%%%
\begin{figure}[t]
\centering
\includegraphics[width=1.0\columnwidth]{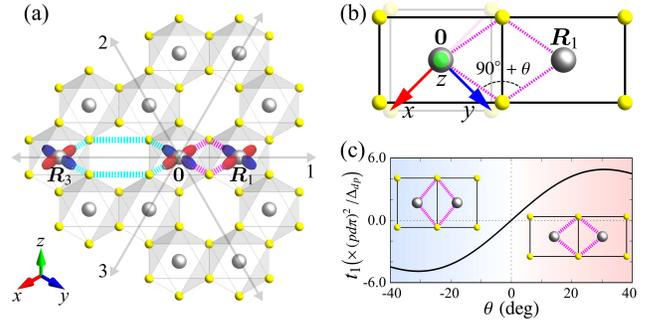}
\caption{
(a)~Schematic picture of a monolayer of honeycomb-layered TM compounds.
The gray and yellow spheres represent the TM cations and the ligand ions, respectively.
The magenta and cyan dotted lines denote indirect paths of nearest- and third-neighbor transfers $t_1$ and $t_3$, respectively, between the $d_{x^2 - y^2}$ orbitals represented by the red and blue ovals.
The arrows 1, 2, and 3 indicate the bond directions in Eq.~\eqref{hr}.  
(b)~Schematic picture of the indirect paths for neighboring octahedra. 
$\theta$ denotes the deviation of the cation-ion-cation angle from $90^{\circ}$, which is caused by a trigonal distortion of the octahedra.
The overlapped faint square represents an undistorted case with $\theta=0^{\circ}$. 
(c)~$\theta$ dependence of the nearest-neighbor transfer $t_1$ in unit of $\left( pd\pi\right)^2/\Delta_{dp}$;  see Eq.~\eqref{t1} and the text. 
Schematic images of distorted octahedra are shown in the inset.
}
\label{setup}
\end{figure}

\begin{figure*}[t]
\centering
\includegraphics[width=2.0\columnwidth]{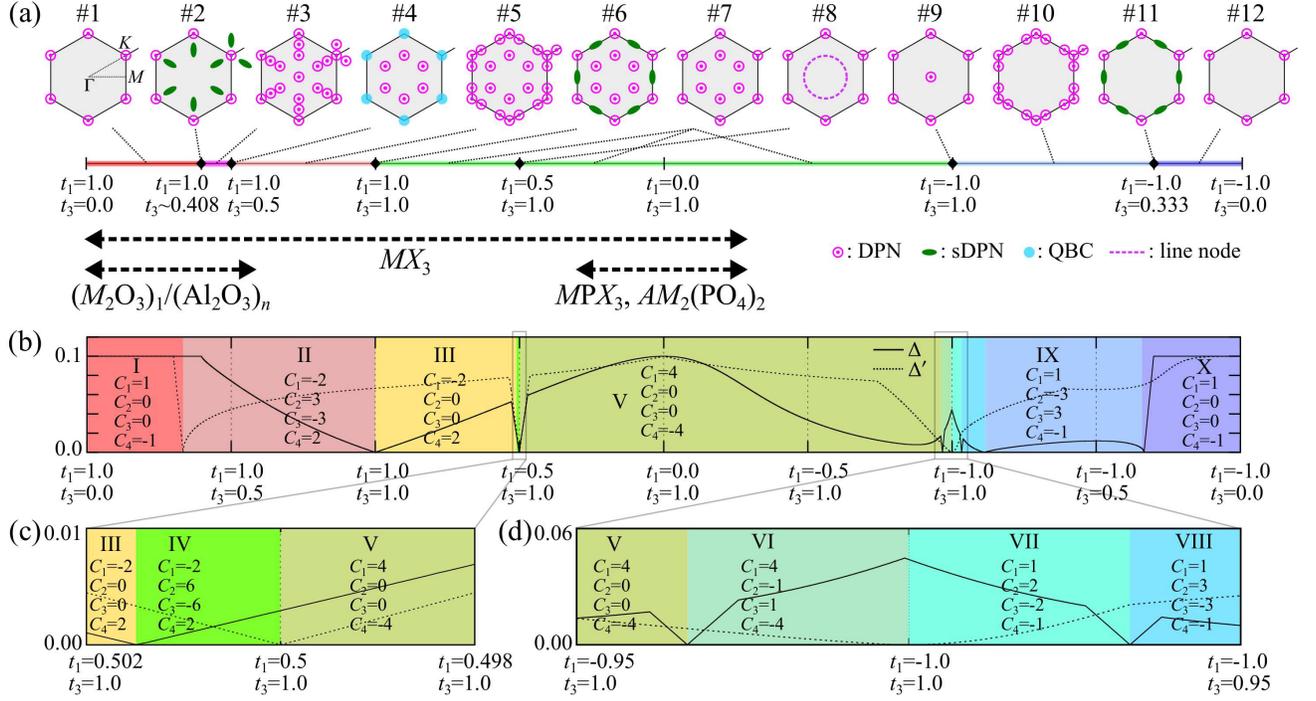}
\caption{
(a)~Phase diagram of the $e_g$-orbital tight-binding model without the SOC at half filling. 
There are 12 states categorized by the number, position, and form of band crossings, which are schematically depicted in momentum space (the gray hexagons represent the first Brillouin zone). 
The results remain the same by changing the signs of $t_1$ and $t_3$ simultaneously. 
Candidate materials are also listed (see the main text). 
(b)~The minimum direct gap opened by the effective SOC $\tilde{\lambda}=0.1$ in Eq.~(\ref{SOC}); $\Delta$ (solid curve) is the value between the second- and third-lowest energy bands, while $\Delta'$ (dotted curve) between the first and second (common to third-fourth).
We also represent the values of the spin Chern number $C_n$ for the $n$th-lowest band, which distinguish 10 states labeled by I-X.
(c) and (d) show the magnified pictures of Fig.~\ref{phase}(b) near the region with $t_1=0.5$ and $t_3=1.0$ and the region with $t_1=-1.0$ and $t_3=1.0$, respectively.  
}
\label{phase}
\end{figure*}

We consider a tight-binding model for the $e_g$-orbital electrons on the honeycomb network of edge-sharing octahedra as shown in Fig.~\ref{setup}(a).
In this structure, the important contributions to the transfer integrals come from the indirect paths via the ligand $p$ orbitals, as the wave functions of the $e_g$ orbitals have large amplitudes in the ligand directions.
We take into account the two types of dominant transfer integrals between the same $e_g$ orbitals for nearest and third neighbors, $t_1$ and $t_3$, as shown by the magenta and cyan lines in Fig.~\ref{setup}(a), respectively, and construct a tight-binding Hamiltonian compatible with the trigonal symmetry of the honeycomb lattice.
The Hamiltonian is given as 
\begin{align}
H =
t_1 \sum_{\left< ij \right>mn\sigma}
\left( \hat{\gamma}_{\alpha_{ij}} \right)_{mn}c^{\dagger}_{im\sigma}c_{jn\sigma} 
+ 
t_3 \sum_{\left< ij \right>^\prime mn\sigma}
\left( \hat{\gamma}_{\alpha_{ij}} \right)_{mn}c^{\dagger}_{im\sigma}c_{jn\sigma},
\label{hr}
\end{align}
where $c^{\dagger}_{i m \sigma}$($c_{i m \sigma}$) is the creation (annihilation) operator of an electron for site $i$, orbital $m = d_{3z^{2}-r^{2}}$ or $d_{x^{2}-y^{2}}$, spin $\sigma =\uparrow$ or $\downarrow$; $\left< ij \right>$ ($\left< ij \right>^\prime$) denotes the nearest (third) neighbors, and $\alpha_{ij} =1$, $2$, or $3$ denotes the bond direction between the sites $i$ and $j$ in Fig.~\ref{setup}(a).
The matrices $\hat{\gamma}_{\alpha_{ij}}$ are obtained as  
$\hat{\gamma}_1
=
\left(
\begin{smallmatrix}
0	&0\\
0	&1\\
\end{smallmatrix}
\right)$, 
$\hat{\gamma}_2 = \hat{\Theta} \hat{\gamma}_1 \hat{\Theta}^{-1}$, and $\hat{\gamma}_3 = \hat{\Theta} \hat{\gamma}_2 \hat{\Theta}^{-1}$, where $\hat{\Theta}$ is the threefold rotational operation in the $e_g$ manifold. 
We will discuss the validity of this model by {\it ab initio} calculations later. 
We note that the $t_1$-$t_3$ model is particle-hole symmetric.

An interesting aspect in this model is that the transfer integrals are sensitively modulated by distortions of the octahedra. 
Let us demonstrate this by considering $t_1$ for a trigonal distortion by compression or expansion of the octahedra perpendicular to the honeycomb plane as shown in Fig.~\ref{setup}(b). 
The modulation of the $d$-$p$-$d$ transfer integral is approximately given as
\begin{equation}
t_{1} 
= 
\frac{
-4\left( pd \pi \right)^2 
+ 
\left(-2 \left( pd \pi \right) + \sqrt{3} \left( pd \sigma \right) \cos^2 \theta \right)^2
}{2\Delta_{dp}}
\sin\theta,
\label{t1}
\end{equation}
where $\left( pd\pi \right)$ and $\left( pd\sigma \right)$ are the Slater-Koster parameters~\cite{PhysRev.94.1498} and $\Delta_{dp} \left(>0\right)$ 
is the energy level splitting between $d$ orbitals of TM cations and $p$ orbitals of ligands.
This relation shows that not only the magnitude but also the sign of $t_1$ is changed by $\theta$. 
Assuming $\left( pd \sigma \right) = -2.2\left( pd\pi\right)$~\cite{harrison2012electronic}, we plot $t_1$ as a function of $\theta$ in Fig.~\ref{setup}(c). 
Note that $t_1$ vanishes in the ideal octahedral case with $\theta=0$.
Considering this aspect, we investigate the electronic structure by taking $t_1$ and $t_3$ as free parameters in the following analysis. 

Figure~\ref{phase}(a) displays the phase diagram for the model (\ref{hr}).
We find 12 states distinguished by the number, position, and form of band crossings at half filling (two electrons per site on average). 
The representative band structures in each state are shown in Supplemental Material~\cite{SM,note}. 
The band crossings evolve systematically while changing $t_1$ and $t_3$. 
For instance, from the state \#1 to \#3, peculiar band crossings appear on the $\Gamma$-$K$ lines at $t_3/t_1 \simeq 0.408$ in addition to the two DPNs at the $K$ points ($K$ and $K'$), and they split into two DPNs each by increasing $t_3$. 
The peculiar band crossing at the state \#2 is quadratic along the $\Gamma$-$K$ lines but linear along the perpendicular directions; we call this type the semi-Dirac point node (sDPN) following the previous works~\cite{PhysRevLett.102.166803,PhysRevLett.103.016402}.
On the other hand, from the state \#3 to \#7, four DPNs at and around the $K$ point merge into a QBC, which is described by the standard effective Hamiltonian~\cite{SM}, and they split again into four; two of them merge again into an sDPN at the $M$ point and finally disappear (gapped out). 
Similar changes are seen from the state \#10 to \#12. 
In the state \#8 at $t_1 / t_3 =0.5$, the six DPNs on the $\Gamma$-$K$ lines are interconnected to form a line node enclosing the $\Gamma$ point. 
Meanwhile, in the state \#9 at $t_1/t_3=-1$, the four DPNs merge with the DPN at the $K$ point, and at the same time, the upper and lower bands meet at the $\Gamma$ point to form a new DPN; the eigenstate of each DPN at the $\Gamma$ or $K$ point is eightfold degenerate, (spin $2$) $\times$ (orbital $2$) $\times$ (sublattice $2$). 
We note that the six DPNs in the state \#7 originate in the ``band folding" by the dominant $t_3$; suppose $t_1=0$, as the lattice sites connected by $t_3$ form a honeycomb structure with the twice larger lattice constant, the DPNs at the $K$ points are copied to the midpoints of $\Gamma$-$K$ lines, as discussed in the previous study~\cite{PhysRevB.97.035125}.

%%%%% SOC %%%%%
The peculiar band crossings that we found can host topologically nontrivial states in the presence of the SOC. 
Although the orbital angular momentum is quenched in the $e_g$ manifold, the $e_g$ manifold is influenced by the SOC through the $e_g$-$t_{2g}$ mixing in distorted octahedra.
In particular, under a trigonal distortion, the leading contribution is given as~\cite{xiao2011interface} 
\begin{equation} 
H_{\rm SOC}
=
-\frac{\tilde{\lambda}}{2}
\sum_{i}\sum_{mn}\sum_{\sigma\sigma'}
c^{\dagger}_{im\sigma}(\hat{\tau}_{y})_{mn}(\hat{\sigma}_{z})_{\sigma\sigma'} c_{in\sigma'},
\label{SOC}
\end{equation}
where $\hat{\tau}_{y}$ ($\hat{\sigma}_{z}$) is the Pauli matrix in orbital (spin) space; here, the $xyz$-axes are taken as shown in the inset of Fig.~\ref{setup}(a) and the quantization axis of spin is taken along the [111] direction. 
The coupling constant is given as $\tilde{\lambda}=\lambda^{2}\Delta_{\rm tri}/\Delta^{2}_{\rm cub}$, where $\lambda$ is the atomic SOC for the $d$ orbitals, $\Delta_{\rm tri}$ is the trigonal field splitting of the $t_{2g}$ orbitals, and $\Delta_{\rm cub}$ is the $e_g$-$t_{2g}$ splitting under the cubic crystal field.

In the presence of the effective SOC in Eq.~(\ref{SOC}), the electronic bands are split into four (twofold degenerate each).
The band splitting is shown by plotting the minimum direct gap between the adjacent bands in Fig.~\ref{phase}(b); 
$\Delta$ denotes the value for the second- and third-lowest bands, while $\Delta'$ for the first and second (common to third-fourth because the energy bands are symmetric with respect to zero energy). 
We also calculate the spin Chern number $C_n$ for the $n$th-lowest band to characterize the topological nature of each gapped state~\cite{doi:10.1143/JPSJ.74.1674}. 
Note that $C_n$ is well defined as the effective SOC does not mix the different spin bands in the present case.

As shown in Figs.~\ref{phase}(b)-\ref{phase}(d), we find ten gapped states I-X with the distinct spin Chern numbers in the presence of the effective SOC in Eq.~(\ref{SOC}).      
Between the gapped states, a band crossing occurs (i.e., $\Delta$ or $\Delta'$ vanishes), and the spin Chern numbers for the crossed bands change their values.
Around $t_1/t_3=0.5$ and $t_1/t_3=-1$, where the system realizes the state \#8 and \#9 in the absence of the SOC, respectively, $\Delta$ and $\Delta'$ change in a complicated manner, as shown in the enlarged figures in Figs.~\ref{phase}(c) and \ref{phase}(d).
Our results indicate that the model (\ref{hr}) exhibits various types of topological transitions in the presence of the SOC.

Interestingly, the spin Chern numbers take unusually high magnitudes in some phases.  
This is conspicuous in the regions where $t_3$ is dominant, e.g., $C=\pm6$ in the state IV and $C=\pm 4$ in the states V and VI. 
The high spin Chern numbers can be traced back to the increased number of band crossing points due to the band folding by $t_3$ in the absence of the SOC [Fig.~\ref{phase}(a)]; the folded bands contribute to enhance the Berry curvatures~\cite{PhysRevB.97.035125}.
Such a folding effect is seen, for instance, in $C_1$ and $C_4$ in the state V that are four times larger than those in the states I and X; this is exactly shown at $t_1 = 0$ where the folded band perfectly overlaps with the original one.
Thus, our results indicate that the honeycomb materials with a substantial contribution from $t_3$ potentially realize high topological numbers.  
The high topological number leads to the corresponding number of edge modes, whose multivalent nature would be useful for a practical application. 

We also note that some bands have odd spin Chern numbers. 
An odd spin Chern number signals a nontrivial state like the $\mathbb{Z}_2$ topological insulator protected by time-reversal symmetry~\cite{PhysRevLett.95.146802}.
Therefore, the states I, VII, and X, where $C_1$ and $C_4$ are odd, share the topological features with the $\mathbb{Z}_2$ topological insulators found in the previous studies for [111] layers of the perovskite structure~\cite{xiao2011interface}.
In addition, our results show that $C_2$ and $C_3$ in the states II and VI and all $C_n$ in the states VI, VIII, and IX are also $\mathbb{Z}_2$ nontrivial.

%%%%% DFT %%%%%
\begin{figure}[t]
\centering
\includegraphics[width=1.0\columnwidth]{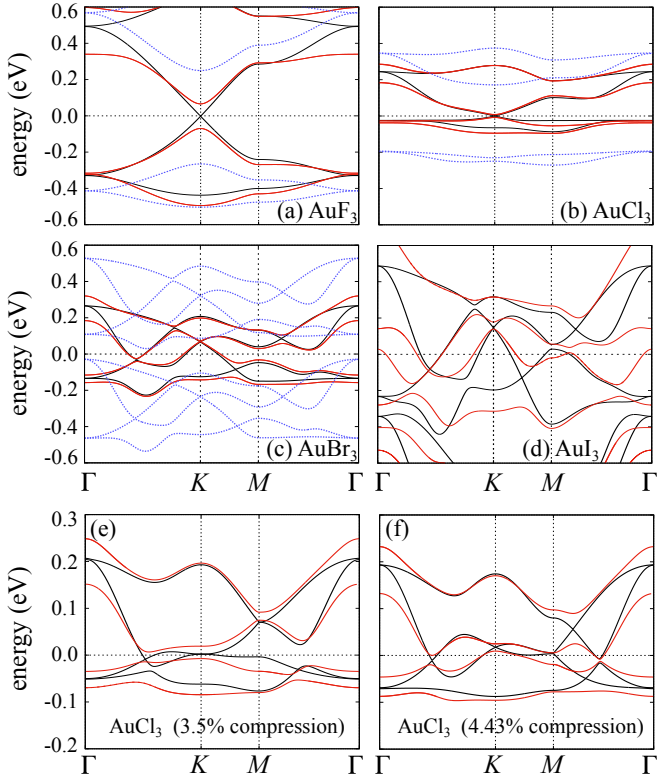}
\caption{
Electronic band structures of monolayers of (a) AuF$_3$, (b) AuCl$_3$, (c) AuBr$_3$, and (d) AuI$_3$.  
The black (red) solid  lines denote the non-relativistic (relativistic) band structures in the paramagnetic state,  which is stable for AuI$_3$ and metastable for others.
The blue dotted lines represent the non-relativistic band structures of the stable magnetic states: the N\`eel antiferromagnetic states for AuF$_3$ and AuCl$_3$ and the ferromagnetic state for AuBr$_3$.  
The Fermi level is set to zero.
(e) and (f) present the paramagnetic band structures for AuCl$_3$ with 3.5\% and 4.43\% compression, respectively.
}
\label{dft}
\end{figure}

Let us further compare our tight-binding analysis with the previous {\it ab initio} studies for the honeycomb-layered TM compounds.
Since our analysis so far is limited to a paramagnetic state, it would apply to weakly correlated materials such as 4$d$ and 5$d$ electron TM compounds.
For instance, for the trichalcogenides $M$P$X_3$ ($M$=Pd, Pt, and $X$=S, Se), multiple DPNs, similar to those in the state \#7, were found around the Fermi level in the paramagnetic state for both 4$d$ Pd and 5$d$ Pt cases, while the former is a metastable state~\cite{PhysRevB.97.035125}.
We note that PdPS$_3$ was synthesized in the bulk form about half a century before, while the electronic state was not studied~\cite{ZAAC:ZAAC19733960305}.
Although for 3$d$-electron systems electron correlations may play a crucial role, a similar band structure was also seen for the $e_g$ manifold in BaFe$_2$(PO$_4$)$_2$ though the Fermi level is in the $t_{2g}$ manifold~\cite{PhysRevB.92.125109,PhysRevB.94.125134,kim2017realizing}.
We note that our analysis also applies to the strongly correlated cases where the electron interaction stabilizes a largely polarized ferromagnetic state and the exchange potential splits the electronic bands into the majority- and minority-spin manifolds.
Indeed, for the trihalide NiCl$_3$~\cite{C6NR08522A} and the layered corundum structure ($M_2$O$_3$)$_1$/(Al$_2$O$_3$)$_5$~\cite{PhysRevB.97.035126}, DPNs similar to those in the state \#1 appear in the spin-polarized $e_g$ manifold.
These observations are summarized below the phase diagram in Fig.~\ref{phase}(a) (for trihalides $MX_3$, see also below). 
Thus, our tight-binding analysis provides a systematic understanding of the existing {\it ab initio} results, and furthermore, a useful guide for further material exploration. 

To confirm our scenario, we discuss the electronic structures of monolayer trihalides, with a focus on the 5$d$ example Au$X_3$ ($X$=F, Cl, Br, and I), based on the {\it ab initio} calculations by \textsc{openmx} code~\cite{openmx} (see Supplemental Material for the computational details~\cite{SM}).
Although the previous experiments for Au$X_3$ reported other crystalline structures in the bulk form~\cite{clark1958crystal,J19670000478,lorcher1975kristallstruktur}, we here consider the honeycomb monolayer form, which is obtained as a locally stable solution by structural optimization in our {\it ab initio} calculations. 
In the non-relativistic calculations, we find that AuI$_3$ is paramagnetic but others are magnetic: the N\`eel antiferromagnetic states for AuF$_3$ and AuCl$_3$ and the ferromagnetic state for AuBr$_3$ [blue dotted lines in Figs.~\ref{dft}(a)-\ref{dft}(c)]~\cite{SM}.
We show the band structures including the paramagnetic solutions (black solid lines) in Figs.~\ref{dft}(a)-\ref{dft}(d). 
Let us discuss the paramagnetic band structures in comparison with our tight-binding results.
We find that AuF$_3$ and AuCl$_3$ possess the DPNs at the $K$ point, similar to the state \#1, while AuBr$_3$ and AuI$_3$ possess the multiple DPNs on the $\Gamma$-$K$ lines, similar to the state \#7 [see Fig.~\ref{phase}(a)].
Furthermore, as shown in Figs.~\ref{dft}(e) and \ref{dft}(f), we find QBCs and sDPNs in AuCl$_3$ with a few percent compression of the lattice structures, which are similar to the states \#4 and \#6, respectively~\cite{SM}.  
Table~\ref{mlwf} summarizes the angle $\theta$ and the transfer integrals estimated by the maximally localized Wannier functions (MLWFs)~\cite{PhysRevB.56.12847,PhysRevB.65.035109} for these cases. 
These results are explained by our tight-binding analysis: the DPNs in AuF$_3$ and AuCl$_3$ are realized by the dominant nearest-neighbor transfer $t_1$ under a substantial compression of the octahedra (large $\theta$), while those in AuBr$_3$ and AuI$_3$ result from the dominant third-neighbor transfer $t_3$ in almost ideal octahedra (small $\theta$); the compressed AuCl$_3$ locates in between.
We confirm that the other transfers, e.g., second- and fourth-neighbor transfers, are less relevant compared to $t_1$ and $t_3$~\cite{SM}, which supports our $t_1$-$t_3$ model.
The results demonstrate the possibility of various band crossings in Fig.~\ref{phase}(a), through the chemical substitution and lattice distortions, once the paramagnetic state is stabilized. 
We note that the band crossings are retained even for the stable ferromagnetic solution for AuBr$_3$ in each exchange-split band, as shown in Fig.~\ref{dft}(c).
When the SOC is included in the relativistic calculations (red dotted lines in Fig.~\ref{dft}), all the DPNs are gapped out, as predicted in our tight-binding analysis. 
We note that the effect of the SOC is relatively large on AuF$_3$ and AuI$_3$, which is also understood by out tight-binding analysis with large $\Delta_{\rm tri}$ and small $\Delta_{\rm cub}$ in Eq.~(\ref{SOC}), respectively. 

\begin{table}[t]
\begin{ruledtabular}
\begin{tabular}{c|cc|cc|cc|cc|cc|cc}
&\multicolumn{2}{c|}{AuF$_3$}  &\multicolumn{2}{c|}{\begin{tabular}{c}AuCl$_3$ \\ (opt.)\end{tabular}}  &\multicolumn{2}{c|}{\begin{tabular}{c}AuCl$_3$ \\ (3.5\%)\end{tabular}}  &\multicolumn{2}{c|}{\begin{tabular}{c}AuCl$_3$ \\ (4.43\%) \end{tabular}}  &\multicolumn{2}{c|}{AuBr$_3$}  &\multicolumn{2}{c}{AuI$_3$}\\
\hline
$\theta$				&\multicolumn{2}{c|}{$17.5^\circ$}	&\multicolumn{2}{c|}{$9.1^\circ$}	&\multicolumn{2}{c|}{$5.1^\circ$}	&\multicolumn{2}{c|}{$3.4^\circ$}	&\multicolumn{2}{c|}{$-0.4^\circ$}	&\multicolumn{2}{c}{$-3.2^\circ$}\\
\hline
$d_{3z^{2}-r^{2}},d_{3z^{2}-r^{2}}$	&-50			&5		&-16			&0				&-2		&-3					&1		&-8					&15		&-22					&53		&-35 \\
$d_{3z^{2}-r^{2}},d_{x^{2}-y^{2}}$	&0			&0		&0			&0				&0		&0					&0		&0					&0		&0					&0		&0	\\
$d_{x^{2}-y^{2}},d_{3z^{2}-r^{2}}$ 	&0			&0		&0			&0				&0		&0					&0		&0					&0		&0					&0		&0	\\
$d_{x^{2}-y^{2}},d_{x^{2}-y^{2}}$	&313			&10		&85			&17				&53		&32					&39		&48					&21		&96					&41		&163	\\
\end{tabular}
\end{ruledtabular}
\caption{
The angle $\theta$ [see Fig.~\ref{setup}(b)] and the transfer integrals between MLWFs obtained by the non-relativistic {\it ab initio} calculations for Au$X_3$ ($X$=F, Cl, Br, and I) with the optimized structures and for AuCl$_3$ with the compressed structures in the paramagnetic solution.
Each value of the transfer integrals means $\bra{d_m,\bm{0}} H \ket{d_n,\bm{r}}$, where $H$ is the Hamiltonian of the system and $\ket{d_m,\bm{r}}$ is the $d_{m}$-like MLWF at site $\bm{r}$ ($m$ = $3z^{2}-r^{2}$ or $x^{2}-y^{2}$).
We take $\bm{r}=\bm{R}_{1}$ and $\bm{R}_{3}$ in the left and right column, respectively [$\bm{0}$, $\bm{R}_{1}$, and $\bm{R}_{3}$ are illustrated in Fig.~\ref{setup}(a)].
The unit of transfer integrals is in meV.
}
\label{mlwf}
\end{table}

%%%%% summary %%%%%
To summarize, we have theoretically investigated the electronic structure of the honeycomb-layered TM compounds with $e_g$ electrons.
We found that the $e_g$ electronic dispersions show a series of peculiar band crossings, and in the presence of the SOC, they turn to be different topological states distinguished by the spin Chern numbers. 
These band crossings and topological states can be realized and controlled by chemical substitutions and distortions of octahedra. 
Indeed, we showed that the previous {\it ab initio} results are understood in a comprehensive manner according to our analysis. Furthermore, we proposed by {\it ab initio} calculations that  trihalides $MX_3$ are good candidates that realize a wide variety of the band topology.
Our findings would stimulate further material exploration toward the exotic phases of matter in the honeycomb-layered materials.
They would also offer a new platform for the study of electron correlation effects on various topological states.

%%%%% acknowledgements %%%%%
\begin{acknowledgments}
Y.S. thanks E.-G. Moon for useful comments.  
The authors acknowledge T. Miyake for helpful comments on the {\it ab initio} calculations.
Y.S. is supported by the Japan Society for the Promotion of Science through a research fellowship for young scientists and the Program for Leading Graduate Schools (MERIT).
\end{acknowledgments}

%%%%% references %%%%%%

\clearpage
\setcounter{figure}{0}
\setcounter{equation}{0}
\setcounter{table}{0}
\renewcommand{\thefigure}{S\arabic{figure}}
\renewcommand{\theequation}{S\arabic{equation}}
\renewcommand{\thetable}{S\arabic{table}}                                                      

\begin{center}                                                    
{\bf ---Supplemental Material---}                               
\end{center}

%%%%% tight-binding bands %%%%% 
\subsection*{S1.~~Electronic band structures of the tight-binding model}
\begin{figure}[b]
\centering
\includegraphics[width=1.0\columnwidth]{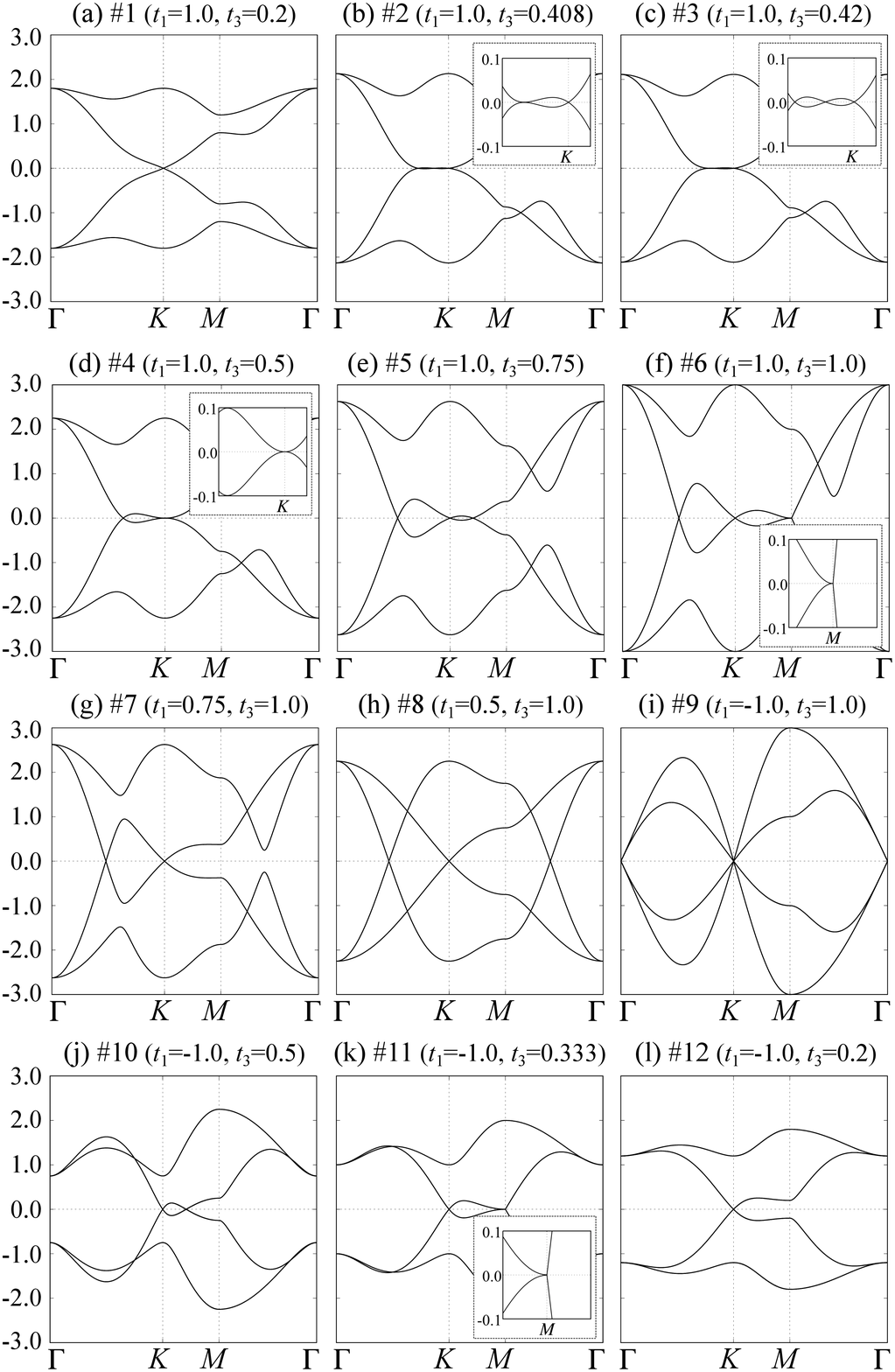}
\caption{Representative electronic band structures in each state in Fig.~2(a) in the main text.
The values of $t_1$ and $t_3$ are denoted in each figure. 
The energy level at half filling is always set to zero.
The insets in (b), (c), and (d) [(f) and (k)] represent the magnified band structures around the $K$($M$) point.
}
\label{band}
\end{figure}
Figure~\ref{band} shows the representative electronic band structures in each state in Fig.~2(a) in the main text.
The band structures are obtained by diagonalizing the tight-binding Hamiltonian in Eq.~(1) in the main text for the values of the nearest- and third-neighbor transfers $t_1$ and $t_3$ indicated in each figure.

We note that, in addition to the half-filled level, the higher and lower two adjacent bands also have crossing points away from the half-filled level. 
For instance, all the band structures possess QBCs at the $\Gamma$ point at nonzero energy, except for the state \#9.
This is closely related to the crystalline symmetry of the honeycomb structure as discussed in the previous theoretical studies on [111] layers of the perovskite structure~\cite{SM:PhysRevB.84.201103,SM:PhysRevB.84.201104,SM:PhysRevB.85.245131}. 
Moreover, several states show DPNs on the $\Gamma$-$K$ and $\Gamma$-$M$ lines at nonzero energy. 
All these additional band crossings are gapped out by the effective SOC and the separated bands acquire nontrivial spin Chern numbers, as discussed in the main text. 
The results indicate that the $e_g$ bands have interesting topology not only at half filling but also for generic fillings.

%%%%% QBC %%%%%                    
\subsection*{S2.~~Effective Hamiltonian for quadratic band crossings}
We derive the effective Hamiltonian for the low-energy excitations in the state \#4 in Fig.~2(a), following the discussion in the previous study~\cite{SM:PhysRevB.84.201103}.
The Fourier transform of the tight-binding Hamiltonian in Eq.~(1) in the main text is written as
\begin{equation}
H
=
\sum_{\bm{k}\sigma}
\begin{pmatrix}
\bm{c}^{\dagger}_{\bm{k} {\rm A} \sigma}	&\bm{c}^{\dagger}_{\bm{k} {\rm B} \sigma}
\end{pmatrix}
\begin{pmatrix}
0								&\hat{H}_{\rm AB}\left(\bm{k}\right)\\
\hat{H}^{\dagger}_{\rm AB}\left(\bm{k}\right)	&0
\end{pmatrix}
\begin{pmatrix}
\bm{c}_{\bm{k} {\rm A} \sigma}\\
\bm{c}_{\bm{k} {\rm B} \sigma}
\end{pmatrix}, 
\label{hk}
\end{equation}
where $\bm{c}^{\dagger}_{\bm{k} \rho \sigma} = (c^{\dagger}_{\bm{k} \rho d_{3z^2-r^2} \sigma} \, \bm{c}^{\dagger}_{\bm{k} \rho d_{x^2-y^2} \sigma})$; $c^{\dagger}_{\bm{k} \rho m \sigma}$ is the creation operator of an electron for wave vector $\bm{k}$, sublattice $\rho=$ A or B, orbital $m = d_{3z^{2}-r^{2}}$ or $d_{x^{2}-y^{2}}$, and spin $\sigma =\uparrow$ or $\downarrow$.
For a small deviation of the wave vector $\bm{k}$ from the $K$ point, the Hamiltonian is rewritten as 
\begin{equation}
\hat{U}^{\dagger}
\begin{pmatrix}
0										&\hat{H}_{AB}\left(\bm{K} + \bm{k}\right)\\
\hat{H}^{\dagger}_{AB}\left(\bm{K} + \bm{k}\right)	&0
\end{pmatrix}
\hat{U}
=
\begin{pmatrix}
\hat{H}_0\left(\bm{k}\right)			&\hat{T}\left(\bm{k}\right)\\	
\hat{T}^{\dagger}\left(\bm{k}\right)	&\hat{H}_1\left(\bm{k}\right)
\end{pmatrix}
+\hat{\mathcal{O}}(k^3),
\label{eq:UHU}
\end{equation}
where $\hat{U}$ is a unitary matrix constructed from the eigenstates at the $K$ point, 
\begin{equation}
\hat{U}
=
\frac{1}{2}
\begin{pmatrix}
\mathrm{i}\sqrt{2}	&0				&-\mathrm{i}	&\mathrm{i}\\	
\sqrt{2}			&0				&1			&-1\\
0				&-\mathrm{i}\sqrt{2}	&\mathrm{i}	&\mathrm{i}\\
0				&\sqrt{2}			&1			&1
\end{pmatrix}.
\end{equation}
We note that $\hat{H}_0\left(\bm{k}\right)$, $\hat{H}_1\left(\bm{k}\right)$, and $\hat{T}\left(\bm{k}\right)$ in Eq.~\eqref{eq:UHU} include $k$-linear terms proportional to $(t_1 - 2 t_3)$.
Thus, when $t_1=2t_3$, we obtain 
\begin{align}
\hat{H}_0\left(\bm{k}\right) &=-\frac{3}{32} t_1 \{ (k^2_x - k^2_y ) \hat{\tau}_x +  2 k_x k_y  \hat{\tau}_y \}, \\ 
\hat{H}_1\left(\bm{k}\right)&=\frac{9}{4}t_1 \{1 - \frac{1}{6}( k^2_x + k^2_y ) \} \hat{\tau_z}, \\ 
\hat{T}\left(\bm{k}\right)&=\hat{\mathcal{O}}(k^2). 
\end{align}
Hence, in the state \#4 where $t_1/t_3=2$, the low-energy excitations are  described by the standard form of the QBCs proportional to $\{ (k^2_x - k^2_y ) \hat{\tau}_x +  2 k_x k_y  \hat{\tau}_y \}$~\cite{SM:PhysRevLett.103.046811}.

%%%%% computational details  %%%%%
\subsection*{S3.~~Computational details of {\it ab initio} calculations}
In the {\it ab initio} calculations in the main text, we used the \textsc{openmx} code~\cite{SM:openmx}, which is based on a linear combination of pseudoatomic orbital formalism~\cite{SM:PhysRevB.67.155108,SM:PhysRevB.69.195113}.
We used the Perdew-Burke-Ernzerhof generalized gradient approximation (GGA) functional in the density functional theory~\cite{SM:PhysRevLett.77.3865}, a $30\times30\times1$ $\bm{k}$-point mesh for the calculations of the self-consistent electron density and the structure relaxation, and vacuum space greater than 10~\AA~between monolayers.
We fully optimized the primitive vectors and atomic positions in the unit cell with the convergence criterion 0.01 eV/\AA~for the inter-atomic forces starting from the ideal honeycomb-crystalline structure in the non-relativistic {\it ab initio} calculations for the paramagnetic solutions; we confirmed that all the crystal structures converge on the honeycomb structures.
We used the optimized structures in the non-relativistic {\it ab initio} calculations for magnetic solutions and the relativistic {\it ab initio} calculations.
In the calculations for AuCl$_3$ with a compressed lattice structure, we reduced the lattice constants uniformly from the optimized values for AlCl$_3$ and relaxed the atomic positions in the unit cell.
To evaluate the transfer integrals, we constructed MLWFs~\cite{SM:PhysRevB.56.12847,SM:PhysRevB.65.035109} via a code implemented in \textsc{openmx}, for the $e_g$ bands obtained by the non-relativistic calculations.

%%%%% magnetism %%%%%
\subsection*{S4.~~Total energy comparison by GGA calculations}
We show the results of non-relativistic GGA calculations for Au$X_3$ ($X$=F, Cl, Br, and I) including magnetic solutions within the primitive cell shown in Fig.~\ref{unit}, paramagnetic (PM), ferromagnetic (FM), and N\'{e}el antiferromagnetic (AFM) states. 
We summarize the energy comparison for these states in Table~\ref{gga}.
The lowest energy solution for $X$=F is AFM though the PM solution is obtained as a metastable state.
In the cases of $X$=Cl and Br [see Figs.~3(b) and 3(c) in the main text], the PM solution remains metastable but has higher energies compared to the magnetic solutions.
The stable state for $X$= Cl (Br) is AFM (FM), whereas the magnetic solutions energetically compete with each other. 
In the case of $X$=I, we obtain only the PM state as the stable solution.

\begin{table}[!h]
\begin{ruledtabular}
\begin{tabular}{l|ccc}
			&PM 	&FM		&AFM\\
\hline
AuF$_{3}$	&$52$	&--	 	&$0$\\
AuCl$_{3}$	&$168$	&$7$ 	&$0$\\
AuBr$_{3}	$	&$91$	&$0$ 	&$10$\\
AuI$_{3}$		&$0$		&-- 		&--
\end{tabular}
\end{ruledtabular}
\caption{
The total energy of each electronic state obtained by GGA calculations.
The lowest energy of all electronic states is set to be zero for each compound.
The blanks indicate that the corresponding state is not obtained as a stable solution.
The unit of energy is in meV per primitive cell (see Fig.~\ref{unit}).
}
\label{gga}
\end{table}

%%%%% Electron transfers %%%%%
\subsection*{S5.~~Transfer integrals of candidate materials obtained by {\it ab initio} calculations}
\begin{figure}[t]
\centering
\includegraphics[width=0.8\columnwidth]{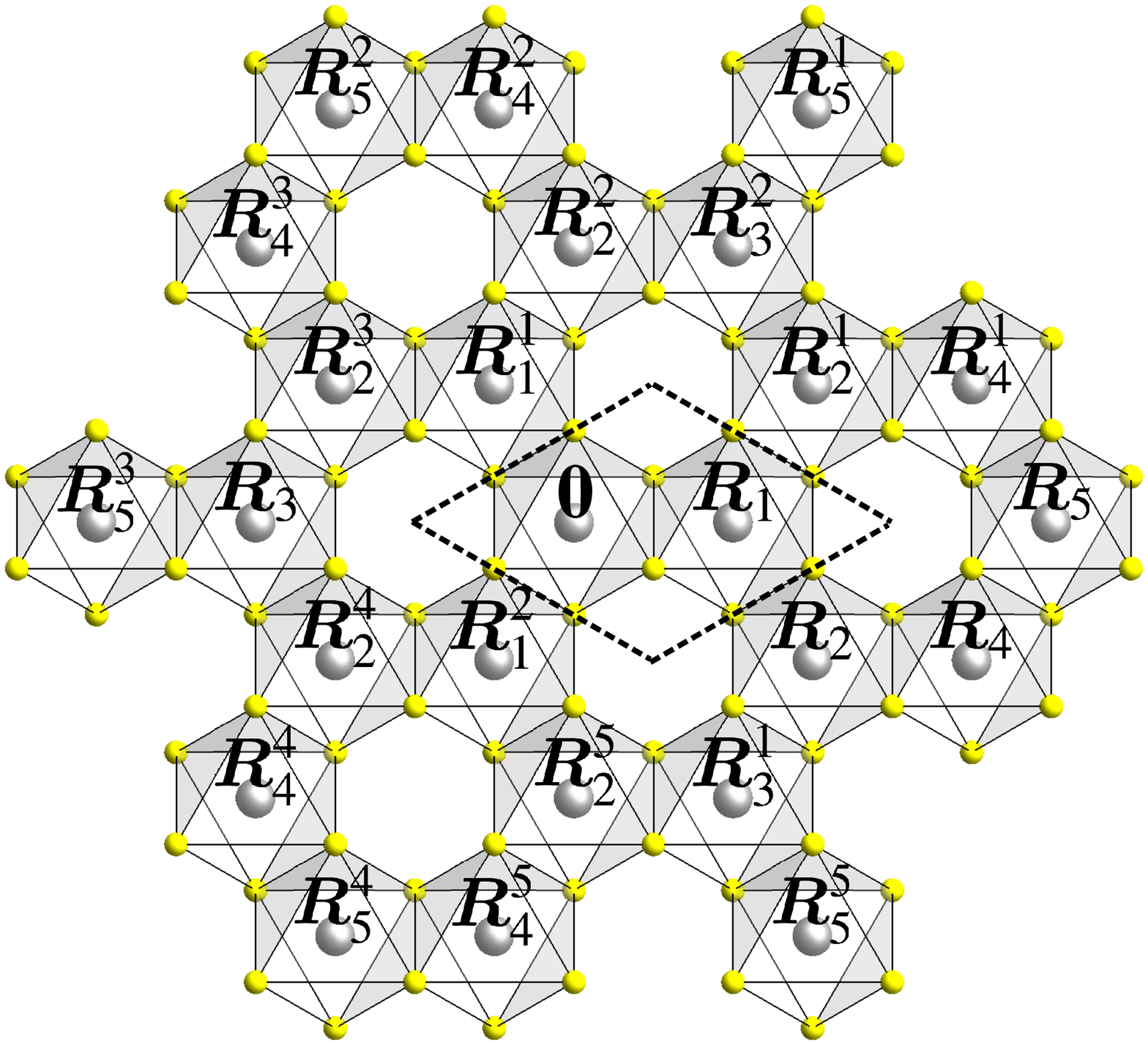}
\caption{
Atomic positions used in the calculation of transfer integrals in Table~\ref{all}.
The dotted lines indicate a regular primitive cell of the honeycomb structure.
}
\label{unit}
\end{figure}

We here provide the detailed information on the transfer integrals estimated by the non-relativistic {\it ab initio} calculations for monolayer honeycomb-layered materials Au$X_3$ ($X$=F, Cl, Br, and I).
In addition to the representative nearest- and third-neighbor transfers in Table~1 in the main text, we list all the transfers in Table~\ref{all} up to fifth neighbors (the spatial positions of the neighbors are illustrated in Fig.~\ref{unit}).
We note that transfer matrices in Table~\ref{all} are related to each other via the threefold rotational operation in the $e_g$ manifold (see the main text).
The results show that the transfer integrals besides for nearest and third neighbors are less relevant, which supports our $t_1$-$t_3$ analysis in the main text.

\begin{table*}[h]
\begin{ruledtabular}
\begin{tabular}{l|ccc|cccccc|ccc}
AuF$_3$					&$\bm{R}_{1}$	&$\bm{R}^{1}_{1}$	&$\bm{R}^{2}_{1}$	&$\bm{R}_{2}$	&$\bm{R}^{1}_{2}$	&$\bm{R}^{2}_{2}$	&$\bm{R}^{3}_{2}$	&$\bm{R}^{4}_{2}$	&$\bm{R}^{5}_{2}$	&$\bm{R}_{3}$	&$\bm{R}^{1}_{3}$	&$\bm{R}^{2}_{3}$\\
\hline
$d_{3z^{2}-r^{2}},d_{3z^{2}-r^{2}}$	&$-50$	&$224$	&$224$			&$0$		&$0$		&$10$	&$0$		&$0$		&$10$			&$5$		&$8$		&$8$\\
$d_{3z^{2}-r^{2}},d_{x^{2}-y^{2}}$	&$0$		&$158$	&$-158$			&$-4$	&$4$		&$-11$	&$17$	&$-17$	&$11$			&$0$		&$2$		&$-2$\\
$d_{x^{2}-y^{2}},d_{3z^{2}-r^{2}}$	&$0$		&$158$	&$-158$			&$17$	&$-17$	&$11$	&$-4$	&$4$		&$-11$			&$0$		&$2$		&$-2$\\
$d_{x^{2}-y^{2}},d_{x^{2}-y^{2}}$	&$313$	&$41$	&$41$			&$7$		&$7$		&$-4$	&$7$		&$7$		&$-4$			&$10$	&$6$		&$6$\\

\hline
						&$\bm{R}_{4}$	&$\bm{R}^{1}_{4}$	&$\bm{R}^{2}_{4}$	&$\bm{R}^{3}_{4}$	&$\bm{R}^{4}_{4}$	&$\bm{R}^{5}_{4}$	&$\bm{R}_{5}$	&$\bm{R}^{1}_{5}$	&$\bm{R}^{2}_{5}$	&$\bm{R}^{3}_{5}$	&$\bm{R}^{4}_{5}$	&$\bm{R}^{5}_{5}$\\
\hline
$d_{3z^{2}-r^{2}},d_{3z^{2}-r^{2}}$	&$-1$	&$-1$	&$0$		&$-2$	&$-2$	&$0$			&$0$		&$1$		&$1$		&$0$		&$1$		&$1$\\
$d_{3z^{2}-r^{2}},d_{x^{2}-y^{2}}$	&$1$		&$-1$	&$-1$	&$0$		&$0$		&$1$			&$0$		&$0$		&$0$		&$0$		&$0$		&$0$\\
$d_{x^{2}-y^{2}},d_{3z^{2}-r^{2}}$	&$1$		&$-1$	&$-1$	&$0$		&$0$		&$1$			&$0$		&$0$		&$0$		&$0$		&$0$		&$0$\\
$d_{x^{2}-y^{2}},d_{x^{2}-y^{2}}$	&$-2$	&$-2$	&$-2$	&$0$		&$0$		&$-2$		&$1$		&$1$		&$1$		&$1$		&$1$		&$1$\\

\hline
\hline
AuCl$_3$ (opt.)				&$\bm{R}_{1}$	&$\bm{R}^{1}_{1}$	&$\bm{R}^{2}_{1}$	&$\bm{R}_{2}$	&$\bm{R}^{1}_{2}$	&$\bm{R}^{2}_{2}$	&$\bm{R}^{3}_{2}$	&$\bm{R}^{4}_{2}$	&$\bm{R}^{5}_{2}$	&$\bm{R}_{3}$	&$\bm{R}^{1}_{3}$	&$\bm{R}^{2}_{3}$\\
\hline
$d_{3z^{2}-r^{2}},d_{3z^{2}-r^{2}}$	&$-16$	&$60$	&$60$			&$0$		&$0$		&$18$	&$0$		&$0$		&$18$			&$0$		&$13$	&$13$\\
$d_{3z^{2}-r^{2}},d_{x^{2}-y^{2}}$	&$0$		&$44$	&$-44$			&$0$		&$0$		&$-10$	&$21$	&$-21$	&$10$			&$0$		&$7$		&$-7$\\
$d_{x^{2}-y^{2}},d_{3z^{2}-r^{2}}$	&$0$		&$44$	&$-44$			&$21$	&$-21$	&$10$	&$0$	&	$4$		&$-10$			&$0$		&$7$		&$-7$\\
$d_{x^{2}-y^{2}},d_{x^{2}-y^{2}}$	&$85$	&$9$		&$9$				&$12$	&$12$	&$-6$	&$12$	&$7$		&$-6$			&$17$	&$4$		&$4$\\

\hline
						&$\bm{R}_{4}$	&$\bm{R}^{1}_{4}$	&$\bm{R}^{2}_{4}$	&$\bm{R}^{3}_{4}$	&$\bm{R}^{4}_{4}$	&$\bm{R}^{5}_{4}$	&$\bm{R}_{5}$	&$\bm{R}^{1}_{5}$	&$\bm{R}^{2}_{5}$	&$\bm{R}^{3}_{5}$	&$\bm{R}^{4}_{5}$	&$\bm{R}^{5}_{5}$\\
\hline
$d_{3z^{2}-r^{2}},d_{3z^{2}-r^{2}}$	&$0$		&$0$		&$2$		&$0$		&$0$		&$2$			&$0$		&$3$		&$3$		&$0$		&$3$		&$3$\\
$d_{3z^{2}-r^{2}},d_{x^{2}-y^{2}}$	&$1$		&$-1$	&$0$		&$1$		&$-1$	&$0$			&$0$		&$-2$	&$2$		&$0$		&$-2$	&$2$\\
$d_{x^{2}-y^{2}},d_{3z^{2}-r^{2}}$	&$1$		&$-1$	&$0$		&$1$		&$-1$	&$0$			&$0$		&$-2$	&$2$		&$0$		&$-2$	&$2$\\
$d_{x^{2}-y^{2}},d_{x^{2}-y^{2}}$	&$1$		&$1$		&$-1$	&$1$		&$1$		&$-1$		&$4$		&$1$		&$1$		&$4$		&$1$		&$1$\\

\hline
\hline
AuCl$_3$ (3.5\%)				&$\bm{R}_{1}$	&$\bm{R}^{1}_{1}$	&$\bm{R}^{2}_{1}$	&$\bm{R}_{2}$	&$\bm{R}^{1}_{2}$	&$\bm{R}^{2}_{2}$	&$\bm{R}^{3}_{2}$	&$\bm{R}^{4}_{2}$	&$\bm{R}^{5}_{2}$	&$\bm{R}_{3}$	&$\bm{R}^{1}_{3}$	&$\bm{R}^{2}_{3}$\\
\hline
$d_{3z^{2}-r^{2}},d_{3z^{2}-r^{2}}$	&$-2$	&$39$	&$39$			&$0$		&$0$		&$13$	&$0$		&$0$		&$13$			&$-3$	&$23$	&$23$\\
$d_{3z^{2}-r^{2}},d_{x^{2}-y^{2}}$	&$0$		&$23$	&$-23$			&$1$		&$-1$	&$-6$	&$14$	&$-14$	&$6$				&$0$		&$15$	&$-15$\\
$d_{x^{2}-y^{2}},d_{3z^{2}-r^{2}}$	&$0$		&$23$	&$-23$			&$14$	&$-14$	&$6$		&$1$		&$-1$	&$-6$			&$0$		&$15$	&$-15$\\
$d_{x^{2}-y^{2}},d_{x^{2}-y^{2}}$	&$53$	&$12$	&$12$			&$9$		&$9$		&$-4$	&$9$		&$9$		&$-4$			&$32$	&$5$		&$5$\\

\hline
							&$\bm{R}_{4}$	&$\bm{R}^{1}_{4}$	&$\bm{R}^{2}_{4}$	&$\bm{R}^{3}_{4}$	&$\bm{R}^{4}_{4}$	&$\bm{R}^{5}_{4}$	&$\bm{R}_{5}$	&$\bm{R}^{1}_{5}$	&$\bm{R}^{2}_{5}$	&$\bm{R}^{3}_{5}$	&$\bm{R}^{4}_{5}$	&$\bm{R}^{5}_{5}$\\
\hline
$d_{3z^{2}-r^{2}},d_{3z^{2}-r^{2}}$	&$0$		&$0$		&$3$		&$0$		&$0$		&$3$			&$-1$	&$5$		&$5$		&$-1$	&$5$		&$5$\\
$d_{3z^{2}-r^{2}},d_{x^{2}-y^{2}}$	&$1$		&$-1$	&$0$		&$2$		&$-2$	&$0$			&$0$		&$-3$	&$3$		&$0$		&$-3$	&$3$\\
$d_{x^{2}-y^{2}},d_{3z^{2}-r^{2}}$	&$1$		&$-1$	&$0$		&$2$		&$-2$	&$0$			&$0$		&$-3$	&$3$		&$0$		&$-3$	&$3$\\
$d_{x^{2}-y^{2}},d_{x^{2}-y^{2}}$	&$2$		&$2$		&$-1$	&$1$		&$1$		&$-1$		&$7$		&$1$		&$1$		&$7$		&$1$		&$1$\\

\hline
\hline
AuCl$_3$ (4.43\%)			&$\bm{R}_{1}$	&$\bm{R}^{1}_{1}$	&$\bm{R}^{2}_{1}$	&$\bm{R}_{2}$	&$\bm{R}^{1}_{2}$	&$\bm{R}^{2}_{2}$	&$\bm{R}^{3}_{2}$	&$\bm{R}^{4}_{2}$	&$\bm{R}^{5}_{2}$	&$\bm{R}_{3}$	&$\bm{R}^{1}_{3}$	&$\bm{R}^{2}_{3}$\\
\hline
$d_{3z^{2}-r^{2}},d_{3z^{2}-r^{2}}$	&$1$		&$30$	&$30$			&$0$		&$0$		&$9$		&$0$		&$0$		&$9$			&$-8$	&$34$	&$34$\\
$d_{3z^{2}-r^{2}},d_{x^{2}-y^{2}}$	&$0$		&$17$	&$-17$			&$2$		&$-2$	&$-3$	&$8$		&$-8$	&$3$			&$0$		&$24$	&$-24$\\
$d_{x^{2}-y^{2}},d_{3z^{2}-r^{2}}$	&$0$		&$17$	&$-17$			&$8$		&$-8$	&$3$		&$2$		&$-2$	&$-3$		&$0$		&$24$	&$-24$\\
$d_{x^{2}-y^{2}},d_{x^{2}-y^{2}}$	&$39$	&$11$	&$11$			&$6$		&$6$		&$-2$	&$6$		&$6$		&$-2$		&$48$	&$6$		&$6$\\

\hline
							&$\bm{R}_{4}$	&$\bm{R}^{1}_{4}$	&$\bm{R}^{2}_{4}$	&$\bm{R}^{3}_{4}$	&$\bm{R}^{4}_{4}$	&$\bm{R}^{5}_{4}$	&$\bm{R}_{5}$	&$\bm{R}^{1}_{5}$	&$\bm{R}^{2}_{5}$	&$\bm{R}^{3}_{5}$	&$\bm{R}^{4}_{5}$	&$\bm{R}^{5}_{5}$\\
\hline
$d_{3z^{2}-r^{2}},d_{3z^{2}-r^{2}}$	&$-1$	&$-1$	&$4$		&$1$		&$1$		&$4$			&$-1$	&$7$		&$7$		&$-1$	&$7$		&$7$\\
$d_{3z^{2}-r^{2}},d_{x^{2}-y^{2}}$	&$2$		&$-2$	&$1$		&$2$		&$-2$	&$-1$		&$0$		&$-5$	&$5$		&$0$		&$-5$	&$5$\\
$d_{x^{2}-y^{2}},d_{3z^{2}-r^{2}}$	&$2$		&$-2$	&$1$		&$2$		&$-2$	&$-1$		&$0$		&$-5$	&$5$		&$0$		&$-5$	&$5$\\
$d_{x^{2}-y^{2}},d_{x^{2}-y^{2}}$	&$3$		&$3$		&$-1$	&$2$		&$2$		&$-1$		&$9$		&$1$		&$1$		&$9$		&$1$		&$1$\\

\hline
\hline
AuBr$_3$					&$\bm{R}_{1}$	&$\bm{R}^{1}_{1}$	&$\bm{R}^{2}_{1}$	&$\bm{R}_{2}$	&$\bm{R}^{1}_{2}$	&$\bm{R}^{2}_{2}$	&$\bm{R}^{3}_{2}$	&$\bm{R}^{4}_{2}$	&$\bm{R}^{5}_{2}$	&$\bm{R}_{3}$	&$\bm{R}^{1}_{3}$	&$\bm{R}^{2}_{3}$\\
\hline
$d_{3z^{2}-r^{2}},d_{3z^{2}-r^{2}}$	&$15$	&$21$	&$21$			&$0$		&$0$		&$2$		&$0$		&$0$		&$2$			&$-22$	&$65$	&$65$\\
$d_{3z^{2}-r^{2}},d_{x^{2}-y^{2}}$	&$0$		&$4$		&$-4$			&$2$		&$-2$	&$1$		&$0$		&$0$		&$-1$		&$0$		&$49$	&$-49$\\
$d_{x^{2}-y^{2}},d_{3z^{2}-r^{2}}$	&$0$		&$4$		&$-4$			&$0$		&$0$		&$-1$	&$2$		&$-2$	&$1$			&$0$		&$49$	&$-49$\\
$d_{x^{2}-y^{2}},d_{x^{2}-y^{2}}$	&$21$	&$18$	&$18$			&$1$		&$1$		&$0$		&$1$		&$1$		&$0$			&$96$	&$8$		&$8$\\

\hline
						&$\bm{R}_{4}$	&$\bm{R}^{1}_{4}$	&$\bm{R}^{2}_{4}$	&$\bm{R}^{3}_{4}$	&$\bm{R}^{4}_{4}$	&$\bm{R}^{5}_{4}$	&$\bm{R}_{5}$	&$\bm{R}^{1}_{5}$	&$\bm{R}^{2}_{5}$	&$\bm{R}^{3}_{5}$	&$\bm{R}^{4}_{5}$	&$\bm{R}^{5}_{5}$\\
\hline
$d_{3z^{2}-r^{2}},d_{3z^{2}-r^{2}}$	&$-3$	&$-3$	&$8$		&$5$		&$5$		&$8$				&$-4$	&$15$	&$15$	&$-4$	&$15$	&$15$\\
$d_{3z^{2}-r^{2}},d_{x^{2}-y^{2}}$	&$2$		&$-2$	&$5$		&$7$		&$-7$	&$-5$			&$0$		&$-11$	&$11$	&$0$		&$-11$	&$11$\\
$d_{x^{2}-y^{2}},d_{3z^{2}-r^{2}}$	&$2$		&$-2$	&$5$		&$7$		&$-7$	&$-5$			&$0$		&$-11$	&$11$	&$0$		&$-11$	&$11$\\
$d_{x^{2}-y^{2}},d_{x^{2}-y^{2}}$		&$10$	&$10$	&$-1$	&$2$		&$2$		&$-1$			&$22$	&$2$		&$2$		&$22$	&$2$		&$2$\\

\hline
\hline
AuI$_3$					&$\bm{R}_{1}$	&$\bm{R}^{1}_{1}$	&$\bm{R}^{2}_{1}$	&$\bm{R}_{2}$	&$\bm{R}^{1}_{2}$	&$\bm{R}^{2}_{2}$	&$\bm{R}^{3}_{2}$	&$\bm{R}^{4}_{2}$	&$\bm{R}^{5}_{2}$	&$\bm{R}_{3}$	&$\bm{R}^{1}_{3}$	&$\bm{R}^{2}_{3}$\\
\hline
$d_{3z^{2}-r^{2}},d_{3z^{2}-r^{2}}$	&$53$	&$44$	&$44$			&$4$		&$4$		&$-9$	&$0$		&$4$		&$4$			&$-35$	&$113$	&$113$\\
$d_{3z^{2}-r^{2}},d_{x^{2}-y^{2}}$	&$0$		&$-5$	&$5$				&$-3$	&$3$		&$5$		&$17$	&$-13$	&$13$		&$0$		&$86$	&$-86$\\
$d_{x^{2}-y^{2}},d_{3z^{2}-r^{2}}$	&$0$		&$-5$	&$5$				&$-13$	&$13$	&$-5$	&$-4$	&$-3$	&$3$			&$0$		&$86$	&$-86$\\
$d_{x^{2}-y^{2}},d_{x^{2}-y^{2}}$	&$41$	&$50$	&$50$			&$-4$	&$-4$	&$9$		&$7$		&$-4$	&$-4$		&$163$	&$14$	&$14$\\

\hline
						&$\bm{R}_{4}$	&$\bm{R}^{1}_{4}$	&$\bm{R}^{2}_{4}$	&$\bm{R}^{3}_{4}$	&$\bm{R}^{4}_{4}$	&$\bm{R}^{5}_{4}$	&$\bm{R}_{5}$	&$\bm{R}^{1}_{5}$	&$\bm{R}^{2}_{5}$	&$\bm{R}^{3}_{5}$	&$\bm{R}^{4}_{5}$	&$\bm{R}^{5}_{5}$\\
\hline
$d_{3z^{2}-r^{2}},d_{3z^{2}-r^{2}}$	&$-12$	&$-12$	&$11$	&$15$	&$15$	&$11$			&$-13$	&$30$	&$30$	&$-13$	&$30$	&$30$\\
$d_{3z^{2}-r^{2}},d_{x^{2}-y^{2}}$	&$-3$	&$3$		&$16$	&$14$	&$-14$	&$-16$			&$0$		&$-25$	&$25$	&$0$		&$-25$	&$25$\\
$d_{x^{2}-y^{2}},d_{3z^{2}-r^{2}}$	&$-3$	&$3$		&$16$	&$14$	&$-14$	&$-16$			&$0$		&$-25$	&$25$	&$0$		&$-25$	&$25$\\
$d_{x^{2}-y^{2}},d_{x^{2}-y^{2}}$	&$21$	&$21$	&$-2$	&$-6$	&$-6$	&$-2$			&$44$	&$2$		&$2$		&$44$	&$2$		&$2$\\
\end{tabular}
\end{ruledtabular}

\caption{
Transfer integrals between MLWFs for candidate materials Au$X_3$ ($X$=F, Cl, Br, and I).
Each value in the table means $\bra{d_m,\bm{0}} H \ket{d_n,\bm{r}}$, where $H$ is the Hamiltonian of the system and $\ket{d_m,\bm{r}}$ is the $d_{m}$-like MLWF at site $\bm{r}$ ($m$ = $3z^{2}-r^{2}$ or $x^{2}-y^{2}$).
See Fig.~\ref{unit} for the spatial positions of $\bm{r}$.
The unit of transfer integrals is in meV.
}
\label{all}
\end{table*}

%%%%% references %%%%%


\begin{thebibliography}{99}
\bibitem{note}
Note that peculiar band crossings, such as DPNs and QBCs, are also found apart from half filling.
\end{thebibliography}
\end{document}